\documentstyle[12pt]{article}

\oddsidemargin=\evensidemargin
\input epsf

\begin{document}

\setlength{\unitlength}{1cm}
\setlength{\baselineskip}{.35in}

\def\shiftleft#1{#1\llap{#1\hskip 0.04em}}
\def\shiftdown#1{#1\llap{\lower.04ex\hbox{#1}}}
\def\thick#1{\shiftdown{\shiftleft{#1}}}
\def\b#1{\thick{\hbox{$#1$}}}

 \begin{titlepage}


\renewcommand{\thefootnote}{\fnsymbol{footnote}}

\begin{center}
{\Large Magnetic Moments of the Octet Baryons
in a Chiral Quark Potential
Model\footnote[2]{supported
by the BMBF under contract number 06T\"{u}746(2)
and by the DFG Graduiertenkolleg MU705/3.
}}
\end{center}

\medskip
\begin{center}
{\large Georg Wagner, Alfons J.\ Buchmann, and Amand Faessler}

Institute for Theoretical Physics,
University of T\"ubingen,
Auf der Morgenstelle 14,
D-72076 T\"ubingen, Germany
\end{center}


\vspace{1.0cm}
\centerline{\bf Abstract}

\vspace{0.2cm}
In quark potential models, two--body current contributions to baryon
magnetic moments arise necessarily to satisfy the continuity
equation for the electromagnetic current.
On the other hand,
the na\"{\i}ve additive quark model predicts the experimental
octet magnetic moments to within 5$\%$.
We demonstrate that consistently derived two--body current
contributions to the octet baryon magnetic moments are
individually large, but tend to cancel each other globally.

\vspace{1.4cm}
\begin{tabbing}
 PACS: \quad\quad\= 12.39.Pn \quad\quad\= Potential models \\
 \> 13.40.Gp \>              Electromagnetic Form Factors \\
 \> 13.40.Em \>              Electric and magnetic moments
\end{tabbing}

\vspace{.4cm}
\begin{tabbing}
Keywords: \quad\= Quark Potential Models, Electromagnetic
		  Two--body Currents, \\
          \>      Continuity Equation, Octet Baryon Magnetic Moments and
		  Form Factors, \\
          \>      Chiral Symmetry
\end{tabbing}

\end{titlepage}


While
the importance of two--body currents for electromagnetic properties of
nuclei is clearly established \cite{Rho79}, the effect of two--body
currents inside baryons is not very well understood
(see ref.\cite{Buc91} and refs.\ therein).
In this respect,
the striking agreement of the na\"{\i}ve constituent quark model
predictions for the octet baryon magnetic moments (summing up the magnetic
moments of three free quarks gives results as close as 5$\%$ to the
experimental values \cite{Kar92,Clo79}) is quite surprising,
and probably has hampered a systematic study of exchange currents in the
baryons for some time.

In some previous works two--body exchange current
contributions to electromagnetic observables of the non--strange
baryons \cite{Buc91} and the deuteron \cite{Buc89} have been studied in detail.
The description of the full baryon octet
involves the extension to non--equal quark masses in the Hamiltonian,
the electromagnetic current operators and the baryon wave functions.

\smallskip
In the chiral quark potential model ($\chi$QPM),
a baryon is described as a nonrelativistic
system of three constituent quarks. The quark dynamics
is governed by two--body potentials.
These potentials are motivated by the main properties of QCD
\cite{Rev} and some experimental facts,
namely asymptotic freedom at short distances,
chiral symmetry and its dynamical breaking at intermediate distances,
and confinement at large distances.
The spontaneous breaking of chiral symmetry
is responsible for the constituent quark mass generation, as
well as for the appearence of the pseudoscalar Goldstone pions or the
pseudoscalar
nonet in the case of three quark flavors. In the chiral quark
potential model, this
is modeled by a one--pion exchange potential between constituent quarks
\cite{Shi84}, together with the exchange potential of its chiral
partner, the scalar--isoscalar sigma meson
\cite{Obu90}\footnote{For
a consistent treatment of chiral symmetry breaking
one should include the exchange of the whole pseudoscalar nonet.
In the present work,
we neglect the Kaon and $\eta$--exchanges
due to their higher mass.}.

\medskip
The $\chi$QPM Hamiltonian in the case of three non--equal
quark masses $m_i$ is written as
\begin{equation}
H=\sum_{i=1}^{3} \big( m_i+ {{\bf p}^2_i\over 2m_i} \big)
-{{\bf P}^2\over 2(m_1+m_2+m_3)}
+ \sum_{i<j}^{3} V^{Conf}({\bf r}_i,{\bf r}_j) +
\sum_{i<j}^{3} V^{Res}({\bf r}_i,{\bf r}_j)\, .
\label{eq:ham}\end{equation}

The confinement potential is given by
($\b{\lambda}_i$ are the Gell--Mann SU(3) color matrices)
\begin{equation}
V^{Conf}({\bf r}_i,{\bf r}_j)=
-a_c \,\b{\lambda}_i\cdot \b{\lambda}_j
\, ({\bf r }_i-{\bf r }_j)^2. \qquad\qquad
\label{eq:conf}
\end{equation}
The residual interactions comprise the chiral interactions
at intermediate and the one--gluon
exchange at short range. The
nonrelativistic one--pion and one--sigma exchange potentials are:
\begin{eqnarray}
V^{OPEP}({\bf r}_i,{\bf r}_j) &=& {g_{\pi q}^2\over {4 \pi }}
\, {\Lambda_\pi^2\over {\Lambda_\pi^2-\mu^2}} \, {\b{\tau}_{i}
\cdot \b{\tau}_{j} \over 4m_im_j}
\, \b{\sigma}_{i}\cdot\b{\nabla}_{r}
\b{\sigma}_{j}\cdot\b{\nabla}_{r}
\, \bigl ( {e^{-\mu r}\over r}- {e^{-\Lambda_\pi r}\over r} \bigr ) \, ,
\label{eq:pion} \\
V^{OSEP}({\bf r}_i,{\bf r}_j) &=& -\, {g_{\sigma q}^2\over {4 \pi}}
\, {\Lambda_\sigma^2\over {\Lambda_\sigma^2-m_{\sigma}^2}}
\, \bigl ( {e^{-m_{\sigma} r}\over r}- {e^{-\Lambda_\sigma r}\over r} \bigr )
\, ,
\label{eq:sigma}
\end{eqnarray}
where
$r=\vert {\bf r} \vert = \vert {\bf r}_i-{\bf r}_j \vert$.
Here $\b{\sigma}_i$ and
$\b{\tau}_i$ denote
the spin and isospin of the i-th quark, respectively.
According to the chiral symmetry arguments in ref.\ \cite{Obu90}
the parameters of the $\pi$-- (coupling only to u-- and d--quarks) and
the flavor--blind $\sigma$--meson are simultaneously fixed:
\begin{equation}
{g^2_{\sigma q}\over {4 \pi}}
= {g^2_{\pi q}\over{4 \pi}} =
\bigg( {3\over 5}{m_u\over m_N} \bigg)^{\!\! 2}
{g_{\pi N}^2\over 4\pi } \quad , \quad
m_{\sigma}^2 = (2 m_u)^2+ \mu^2  \quad ,\quad
\Lambda_{\pi} = {\Lambda_{\sigma}} \equiv \Lambda.
\label{eq:cutoffs}
\end{equation}
We use $g_{\pi N}^2/4\pi$=13.845 and the pion mass $\mu$=138 MeV.
The cut--off $\Lambda$ describes the
range of the meson--quark interaction region,
due to the finite sizes of the constituent quark and the mesons.
The one--gluon exchange potential is used in the well known form of
\cite{deR75}:
\begin{eqnarray}
V^{OGEP} ({\bf r}_i,{\bf r}_j) \!\!\!\! &=& \!\!\!\!
{\alpha_{s}\over 4}\b{\lambda}_{i}
\!\cdot\!\b{\lambda}_{j} \!\biggl\lbrace
{1\over r}-{\pi\over 2}({1\over m_i^2}+{1\over m_j^2}+{4\over 3}
{\b{\sigma}_{i}\cdot\b{\sigma}_{j}\over m_im_j} )
\delta({\bf r})
\! -\! {1\over 4m_im_j}
(3\b{\sigma}_i\cdot{\bf{\hat r}}
\b{\sigma}_j\cdot{\bf {\hat r}}
-\b{\sigma}_i\cdot\b{\sigma}_j) {1\over r^3} \nonumber\\
&-& \!\!\!\!
{1\over 8r^3}\!\bigg( 3\big( {\bf r}\!\times\!
\! ( {{\bf p}_i\over m_i}-{{\bf p}_j\over m_j} )\big)\!\cdot\!
( {\b{\sigma}_i\over m_i}+{\b{\sigma}_j\over m_j} )
\! -\! \big( {\bf r}\!\times\!
\! ( {{\bf p}_i\over m_i}+{{\bf p}_j\over m_j} )\big)\!\cdot\!
( {\b{\sigma}_i\over m_i}-
{\b{\sigma}_j\over m_j} ) \bigg) \biggr\rbrace
\; .
\label{eq:gluon}
\end{eqnarray}

For the baryon wave functions, we use simple $(0s)^3$ harmonic oscillator
states.
The harmonic oscillator lengthes $b_{\b{\rho}}$ and
$b_{\b{\lambda}}$ in the directions of the
internal Jacobi--coordinates
$\b{\rho}    = ({\bf r}_1-{\bf r}_2) $ and
$\b{\lambda} = {\bf r}_3+(m_1{\bf r}_1+m_2{\bf r}_2)/(m_1+m_2)$
can be expressed in terms of quark mass ratios and the oscillator length
$b_N$ used for the nucleon and the $\Delta$--resonance.
The orbital part of the wave function
$\Phi_{3q}(f_1f_2f_3,\b{\rho},\b{\lambda})$
depends on the 3 quark flavors $f_1$,$f_2$,$f_3$,
and the full wave function can be written as
\begin{equation}
\vert B \rangle  =
\frac{1}{\sqrt{2}} \!\sum_{S_{12}=0,1}\!\bigg(
\sum_{f_1f_2f_3} \! ^B\! C^{S_{12}}(f_1f_2f_3)\times
\Phi_{3q}(f_1f_2f_3,\b{\rho},\b{\lambda})\times\vert f_1f_2f_3\rangle\bigg)
\times\vert S,S_z;S_{12}\rangle\times \vert\;
                  { \parbox{.4cm}{
                    \begin{picture}(.3,.9)
                      \thicklines
                      \put(0.,0.){\framebox(.3,.3){}}
                      \put(0.,.3){\framebox(.3,.3){}}
                      \put(0.,.6){\framebox(.3,.3){}}
                    \end{picture} }_C } \rangle .
\label{eq:baryon}
\end{equation}
$\vert f_1f_2f_3\rangle$ is the flavor part of the wave function, and the
antisymmetry lies in the color space.
The coefficients $^B\! C^{S_{12}}(f_1f_2f_3)$ are SU(3) Clebsch--Gordan
coefficients. These wave functions can be found for example in
\cite{Clo79}. We use a different overall sign for the mixed
symmetric states than in \cite{Clo79}.

\smallskip
In this form the $\chi$QPM has been used to calculate the nucleon and
$\Delta$ spectrum (positive parity), and has also been
applied to the NN system with
a fair amount of success \cite{Obu90,Buc95,Val94}.
As usual, the parameters for the present calculation
are fitted to the nucleon and $\Delta$--resonance
masses.

\marginpar{table 1}

\bigskip
The baryon masses $m_B$ are determined as the expectation values of
Hamiltonian (\ref{eq:ham}) for the wave functions (\ref{eq:baryon}).
The result is shown in table \ref{table:masses}.

\marginpar{table 2}

\bigskip
Gauge invariance is the key requirement when calculating electromagnetic
properties of the baryons. Any isospin-- or momentum--dependent
two--body potential in the Hamiltonian gives rise to additional
two--body exchange current contributions to the total
electromagnetic current. These two--body currents are necessary
in order to satisfy the continuity equation for the electromagnetic current.
Here, we list the electromagnetic currents, that have to be considered for
consistency with our Hamiltonian of eq.(\ref{eq:ham}).
They have been derived in a number of papers
\cite{Buc91,Buc89} for the case of non--strange quarks.

The nonrelativistic one--body
current of the quarks of figure 1(a),
referred to as impulse approximation, is given by
\begin{equation}
{\bf J}_{imp} ({\bf r}_i,{\bf q)} =
{e_i\over2m_i} \Bigl( i \lbrack \b{\sigma}_i\times
{\bf p}_i,e^{i{\bf{q \cdot r }}_i} \rbrack
+ \lbrace {\bf p}_i, e^{i{\bf{q \cdot r }}_i} \rbrace \Bigr).
\label{eq:impulse}
\end{equation}

\marginpar{figure 1}

\bigskip
In the SU(3)--flavor case the charge operator for pointlike quarks is
\begin{equation}
e_i=\frac{e}{2}\, \bigg( B_i + S_i + \b{\tau}_z^{(i)} \bigg) \quad ,
\label{eq:charge}
\end{equation}
where $B_i$=1/3 is the baryon number and $S_i$ is the strangeness
quantum number of the i--th quark.
Within the picture of
dressed quasi--particle constituent quarks it is
necessary to
introduce a finite electromagnetic size of the quarks \cite{Buc91}.
Motivated by the vector meson dominance picture \cite{Vog90}, a
possible parametrization for the electromagnetic
quark size is to multiply the expression for the
quark charge by a simple monopole form.
The electromagnetic constituent quark size
$r^2_{\gamma q}$=0.36 fm$^2$
(independent of SU(3)--flavor) is related
to the vector--meson mass, and is taken from the
calculation of the charge radii of the nucleons \cite{Buc91}.
We multiply the
expressions for the current operators by the electromagnetic quark
form factor
\begin{equation}
F_{\gamma q}({\bf q}^2) =
\bigg( 1 + { {\bf q}^2 \, r_{\gamma q}^2\over 6} \bigg)^{-1} \qquad .
\label{eq:monopole}
\end{equation}

For the pion, due to its isospin structure, we get the well known
pion--pair ($\pi q\bar q$) current, figure 1(b),
and the isovector pionic ($\gamma\pi\pi$) current,
figure 1(c):
\begin{eqnarray}
{\bf J}^{IS}_{\pi q {\bar q}}({\bf r}_i,{\bf r}_j,{\bf q}) \!\!\! &=&\!\!\!
{ie\over 6} {g^2_{\pi q}\over 4\pi }
{\Lambda^2 \over \Lambda^2-\mu^2}
\bigg\{ {{\b{\tau}_i}\!\cdot\! {\b{\tau}_j} \over 8m_i^3m_j}
\, e^{i{\bf q}\cdot {\bf r}_i} {\bf q}\times {\bf {\nabla_r}}
{\b{\sigma}_j}\cdot{\bf {\nabla_r}}
\bigl ({e^{-\mu r}\over r}- {e^{-\Lambda r}\over r} \bigr )
+(i\leftrightarrow j)\bigg\}
\nonumber \\
{\bf J}^{IV}_{\pi q {\bar q}}({\bf r}_i,{\bf r}_j,{\bf q}) \!\!\! &=&\!\!\!
e {g^2_{\pi q}\over 4\pi }
{\Lambda^2 \over \Lambda^2-\mu^2}
{\Bigl\lbrace
{({\b{\tau}_i}\!\times\! {\b{\tau}_j})_3 \over 4m_im_j}
{e^{i{\bf q}\cdot {\bf r}_i}}\b {\sigma}_i \
{\b{\sigma}_j}\cdot{\bf {\nabla_r}}+(i\leftrightarrow j)\Bigr\rbrace}
\bigl ({e^{-\mu r}\over r}- {e^{-\Lambda r}\over r} \bigr )
\label{eq:pioncurrent} \\
{\bf J}^{IV}_{\gamma \pi \pi}({\bf r}_i,{\bf r}_j,{\bf q}) \!\!\! &=&\!\!\!
e {g^2_{\pi q}\over 4\pi }
{\Lambda^2 \over \Lambda^2-\mu^2}
{({\b{\tau}_i}\!\times\! {\b{\tau}_j})_3 \over 4m_im_j}
\b {\sigma}_i \cdot  {\bf \nabla}_i
\b {\sigma}_j \cdot  {\bf \nabla}_j
\!\int_{-1/2}^{1/2} \!\!\!\!\! dv
e^{i{\bf q}\cdot ({\bf R}-{\bf r}v)}
\bigg({\bf z}_{\mu} { e^{-L_{\mu }r}\over L_{\mu} r}
- {\bf z}_{\Lambda} {e^{-L_{\Lambda }r}\over L_{\Lambda} r} \bigg) .
\nonumber
\end{eqnarray}
For the pionic current
we have used the abbreviations
${\bf R}= ({\bf r}_i+{\bf r}_j)/2  $,
${\bf z}_m({\bf q},{\bf r})=L_m{\bf r}+ivr{\bf q}$,
and
$L_{m}(q,v) = [{1\over 4}q^2(1-4v^2)+m^2]^{1/2} $.
We use the same monopole form factor of eq.(\ref{eq:monopole})
for the extended photon--pion vertex as
for the photon--quark vertex.

\smallskip
The gluon--pair current of figure 1(d) is given by
\begin{equation}
{\bf J}_{gq{\bar q}} ({\bf r}_i,{\bf r}_j,{\bf q})=-
{\alpha_s\over 8}\,{\b{\lambda}_i}\cdot{\b{\lambda}_j}
{\Bigl\lbrace {{e_i\over m_i} \, e^{i{\bf q}\cdot {\bf r}_i}}
({{\b {\sigma}_i}\over m_i}+{{\b {\sigma}_j}\over m_j})\times{\bf r}
+(i\leftrightarrow j)\Bigr\rbrace}
{1\over r^3}. \qquad\quad
\label{eq:gluonpair}
\end{equation}
The gluon--pair current is related to the spin--orbit part in
eq.(\ref{eq:gluon}). As discussed in \cite{Buc91}, one should then for
consistency also consider two--body currents corresponding to the
scalar confinement-- eq.(\ref{eq:conf}) and sigma--potentials
eq.(\ref{eq:sigma}), in the following generically denoted as $V_C^S$.
Including the lowest order ${\cal{O}} \big( {1\over m_q^2}\big)$
relativistic corrections leads to the following
structure of the scalar exchange potential:
\begin{eqnarray}
V^S &=& V_C^S -{1\over 4}\bigg( \!\big\{
( {\b{\sigma}_i\cdot {\bf p}_i\over m_i} )^2,V_C^S\big\}
+ \!\big\{
( {\b{\sigma}_j\cdot {\bf p}_j\over m_j} )^2,V_C^S\big\} +
{1\over 2} ({{\b{\nabla}}_i^2\over m_i^2}
+ {{\b{\nabla}}_j^2\over m_j^2}) V_C^S \nonumber\\
& &\! + {1\over r} {{\rm d} V_C^S\over {\rm d} r}
\bigg[ {1\over 2} ({\b{\sigma}_i\over m_i}+{\b{\sigma}_j\over m_j})\cdot
{\bf r} \times ({{\bf p}_i\over m_i}-{{\bf p}_j\over m_j} ) +
{1\over 2} ({{\b{\sigma}}_i\over m_i}-{{\b{\sigma}}_j\over m_j})
\cdot {\bf r} \times ({{\bf p}_i\over m_i}+{{\bf p}_j\over m_j} )
\bigg] \bigg) .
\label{eq:relativ}
\end{eqnarray}
Also from the baryon mass spectroscopy
and the NN phase shift analysis, these spin--orbit terms in the
scalar potentials seem to be needed \cite{Val94,Isg78}.
Similarily to the gluon case, they provide by
minimal substitution a two--body confinement and
sigma--pair current, see
figure 1(e). This current
can be expressed by use of $V^{Conf}$ of eq.(\ref{eq:conf}) as:
\begin{equation}
{\bf J}_{conf}({\bf r}_i,{\bf r}_j,{\bf q})=
- \Bigl \{  {e_i\over 2m_i^2} e^{i {\bf q}\cdot {\bf r}_i}
\, V^{Conf} \, i \b{\sigma}_i \times {\bf q}
+(i\leftrightarrow j)  \Bigr \} \quad .
\label{eq:jconf}
\end{equation}
By analogy \cite{Buc91}, we have to replace $V^{Conf}$ in (\ref{eq:jconf})
by the potential $V^{OSEP}$ of eq.(\ref{eq:sigma}) to obtain the
sigma--pair current.

\bigskip
Calculating the magnetic form factors involves taking matrix elements of the
above currents. The full form factor can be written as a sum
\begin{equation}
F^B({\bf q}^2) = F^B_{imp} + F^B_{gq\bar q} + F^B_{\pi q\bar q} +
F^B_{\gamma\pi\pi} + F^B_{conf} + F^B_\sigma \quad .
\label{eq:formfactor}
\end{equation}
The different contributions to the magnetic moments
$\mu^B = F^B({\bf q}^2=0)$
for the octet baryons as calculated with the parameters
of table \ref{table:parameters} are given in table \ref{table:mamo}.

\marginpar{table 3}

\bigskip
First we recognize that the individual two--body currents contributions
are rather large, especially the gluon and confinement part.

We find a
slightly better overall agreement of our total result with
the experimental values in comparison to the impulse result alone.
We should stress, that none of the parameters is fitted to the
baryon magnetic moments, but to the nucleon and $\Delta$--masses, and that
these parameters allow to describe also the
static deuteron properties \cite{Buc95} with reasonable accuracy.
By varying the constituent quark masses (for example $m_u$=324 MeV,
$m_u/m_s$=0.6), we could improve our
total result to get the aforementioned 5$\%$ agreement without
giving up the good agreement for the baryon masses.

Because the isovector pion--pair and pionic currents
largely cancel each other (for the proton we find
${{\mu}}^{IV}_{\pi q\bar q}$=--0.29 n.m.\ and
${{\mu}}^{IV}_{\gamma\pi\pi}$=0.41 n.m.), the
pion cloud gives only a small
correction for the nucleon magnetic moments\footnote{We
expect (in addition to a suppression due to the rather big Kaon mass)
the same cancellation of the Kaon--pair current and the "Kaon--in--flight"
contribution. This can be considered as further justification for our
exclusion of Kaon and $\eta$ degrees of freedom in the present calculation.}.
To the charged $\Sigma^\pm$ only the small isoscalar pion--pair
current contributes.

The second cancellation occurs between the gluon--, confinement-- and
$\sigma$--pair currents.
This cancellation is clearly favored for a quark core radius
$b_N \simeq$ 0.6 fm.
This value for $b_N$ is also obtained from the variational principle
${\partial M_N(b_N)\over \partial b_N}=0$
for $(0s)^3$ shell model trial wave functions \cite{Buc91}.
Furthermore, one can establish in this basis a relation between
the neutron charge radius and the nucleon--$\Delta$ mass--splitting
$< r^2 >_n =- b_N^2 {M_\Delta-M_N\over M_N}$, which is fulfilled for
$b_N$=0.612 fm \cite{Buc91}.
The choice of $b_N\simeq$
0.6 fm is therefore the most consistent one in the restricted basis of
$(0s)^3$--states.
Taking for example $b_N$ = 0.5 fm would completely
spoil these cancellations, and we
would find for the proton a magnetic moment of
${\mu}_p\simeq 5.6$ n.m.
Another point to mention at this stage is, that the two--body
current associated with
a vector rather than a scalar confinement has the same sign
as the gluon exchange current for all the octet baryons, i.e.\ the
wrong sign to cancel the large gluon contribution.
 This has been observed before \cite{Buc91} for the
case of the proton and neutron magnetic moments.

\smallskip
Let us now study the magnetic radii.
The magnetic radii of hadrons provide valuable information about the
inner structure of hadrons.
In particular, they provide a further test of the gluon, pion,
and scalar exchange currents discussed here.
In table \ref{table:mara}, we present the magnetic radii
\begin{equation}
\langle r^2\rangle^B =
- {6\over \mu^B}
\, {{\rm d} F^B({\bf q}^2)\over {\rm d} {\bf q}^2}
\bigg\vert_{{\bf q}^2=0} =
\langle r^2\rangle_{imp} + \langle r^2\rangle_{gq\bar q} +
\langle r^2\rangle_{\pi} + \langle r^2\rangle_{conf} +
\langle r^2\rangle_{\sigma} \quad .
\end{equation}

\marginpar{table 4}

\bigskip
First, we recognize, that
the experimental proton and neutron magnetic radii can be nicely reproduced
in our calculation.
Throughout the whole octet, the net contribution of all the exchange
current processes to the magnetic radii is quite small,
however always negative (except for the $\Xi^-$).
 Thus the cancellation of two--body effects
is also seen in the magnetic radii of the octet baryons.
We observe that the strange baryons have a slightly smaller
magnetic size than the proton and neutron.
Unfortunately, there are no experimental data for comparison up to now.
With the present
experimental errors for the nucleon magnetic radii, it is
difficult to find clear--cut evidence for
exchange--current effects in these observables.

\smallskip
In this letter we have investigated the gauge invariant pion, gluon
and scalar exchange current contributions to the magnetic moments and radii
of the octet baryons.
We have found substantial cancellations between different exchange current
contributions.
Because we neglect various other effects, such as small
anomalous magnetic moments of the quarks, configuration mixing,
relativistic effects beyond the two--body currents included here,
one should not expect a perfect agreement with the experimental data.
Nevertheless, we stress again, that our results
are obtained without fitting any parameter to the magnetic moments and
the magnetic radii, but
using parameters fixed by the nucleon and $\Delta$--masses.
It has been shown before, that the aforementioned cancellations
between different exchange currents also occur, when improved
wave functions (configuration mixing) are used \cite{Buc91}.

Naturally the question arises:
Are these cancellations between various two--body currents
a coincidence, or is there a more fundamental reason why the
cancellations are most pronounced just for
$b_N\simeq$ 0.6 fm = ${1\over {\rm 329 MeV}} \simeq {1\over m_u}$?
In the latter case, the baryon magnetic moments could
provide another argument for a universal
nucleon harmonic oscillator length of $b_N\simeq$0.6 fm. This value gives
a reasonable baryon size, and allows to describe the
positive parity baryon spectrum and
baryon electromagnetic observables with the same parameters.

In summary our main result is,
that the --- individually large --- two--body contributions
to the octet baryon magnetic moments tend to cancel each other, when the
exchange currents are consistently evaluated with the Hamiltonian.
Therefore our results offer a possible explanation for the
surprisingly good agreement of the na\"{\i}ve additive quark
model predictions with the experimental
values.

\setlength{\baselineskip}{.22in}

\newpage
{\large\bf Figure captions:}

\vspace{.5cm}
Figure 1:(a) Impulse current, (b) pion--pair current,
(c) pionic current, (d) gluon--pair current, and (e) scalar exchange current
(confinement and $\sigma$--exchange). The black dots on the photon--quark
and the photon--pion vertices
indicate the finite electromagnetic size of the constituent quark and the pion,
for which we take $r_{\gamma q}^2$=$r_{\gamma\pi}^2$=0.36 fm$^2$.

\newpage
{\Large Georg Wagner {\sl et al.}, {\bf Figure 1},

\vspace{.2cm}
"Magnetic Moments of the
Octet ...", Phys.Lett.B}

\vspace{1.5cm}

\begin{figure}[h]
\begin{center}
  {\epsffile{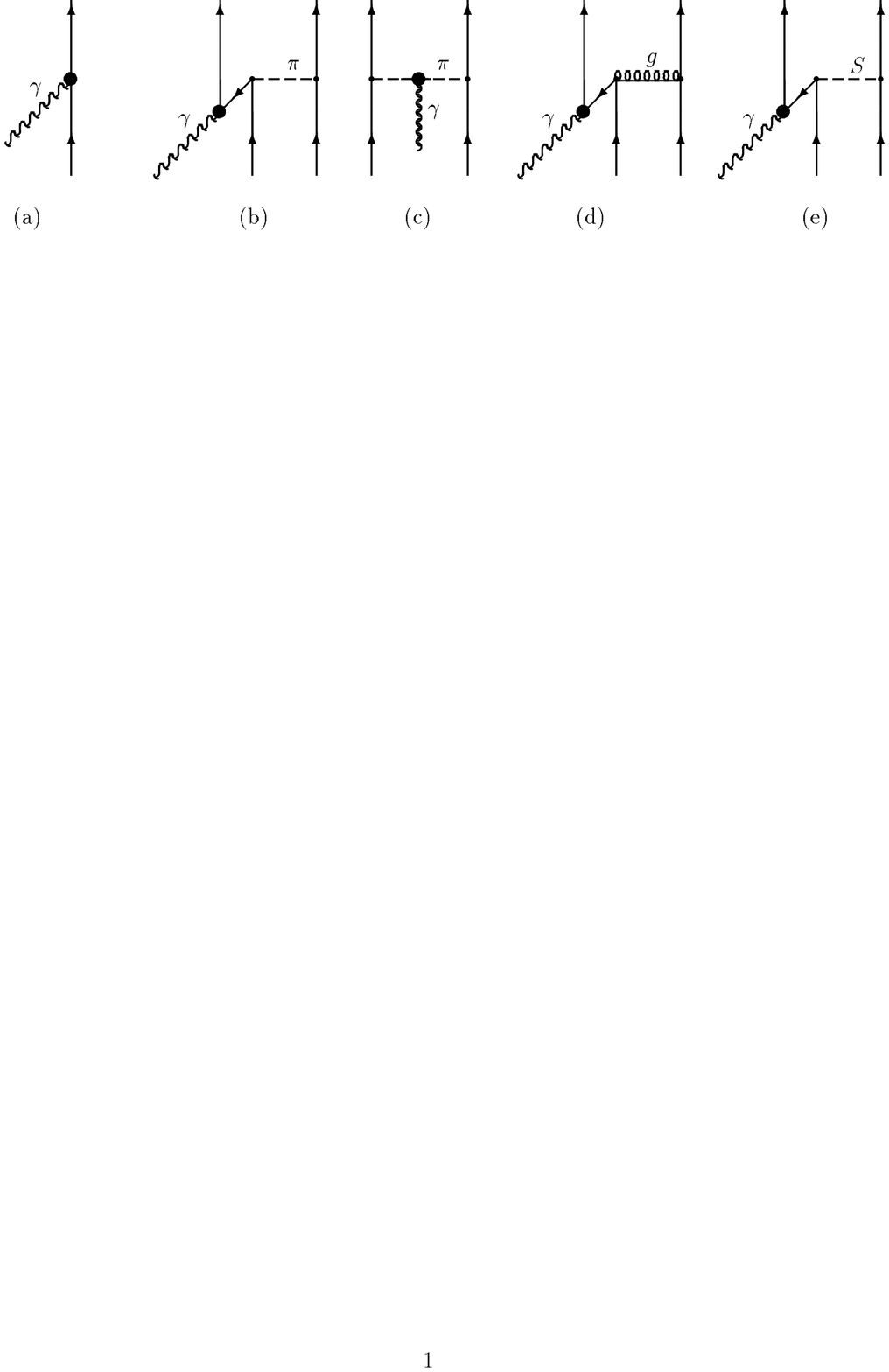}}
\label{figure:currents}
\end{center}
\end{figure}

\newpage
{\large\bf Tables:}

\vspace{.5cm}

\begin{table}[h]
\begin{center}
\begin{tabular}{| l | l | l | l | l | l |} \hline
$b_N$ & $m_u$=$m_d$ & $m_u/m_s$ & $\Lambda$ & $a_c$  &
$\alpha_s$ \\ \hline\hline
0.613 fm & 313 MeV & 0.6 & 4.2 fm$^{-1}$ & 20.92 MeV fm$^{-2}$ & 1.094 \\
\hline
\end{tabular}
\caption{The parameters of Hamiltonian (1) for the present calculation}
\label{table:parameters}
\end{center}
\end{table}

\begin{table}[h]
\begin{center}
\begin{tabular}{| l || r | r | r | r | r | r || r | r |} \hline
 & $\sum_i m_i$ & $E_{kin}$ &
 $V_{conf}$ & $V_{gluon}$ & $V_\pi$ & $V_\sigma$ & $m_B$ &
  $m_{exp}$ \cite{Agu94} \\ \hline\hline
p,n       &   939 & 497 & 189 & -512 & -119 &  -54 &  939 &  939 \\ \hline
$\Delta$  &   939 & 497 & 189 & -314 &  -24 &  -54 & 1232 & 1232 \\ \hline
$\Sigma$  &  1148 & 497 & 163 & -540 &   -8 &  -63 & 1197 & 1193 \\ \hline
$\Lambda$ &  1148 & 497 & 163 & -561 &  -71 &  -63 & 1112 & 1116 \\ \hline
$\Xi$     &  1356 & 497 & 138 & -613 &    0 &  -74 & 1304 & 1318 \\ \hline
\end{tabular}
\caption{The different contributions of Hamiltonian
(1) to the baryon masses. All quantities are given in [MeV]. The experimental
values represent averages over particles with different charge.}
\label{table:masses}
\end{center}
\end{table}

\newpage
\begin{table}[t]
\begin{center}
\begin{tabular}{| l || r | r | r | r | r || r | l |} \hline
octet & ${\mu}_{imp}$ & ${\mu}_{gq\bar q}$ &
  ${\mu}_\pi$ & ${\mu}_{conf}$ &
  ${\mu}_\sigma$ & ${\mu}_{tot}$ & ${\mu}_{exp}$ \cite{Agu94}
\\ \hline\hline
p &  3.000 &  0.598 &  0.149 &  -1.205  &  0.347  &  2.889  &  2.793 \\ \hline
n & -2.000 & -0.199 & -0.098 &   0.804  & -0.232  & -1.725  & -1.913 \\ \hline
$\Sigma^+$  &
 2.867 &  0.615 & 0.017 & -1.003 &  0.363 &  2.859 &  2.458 $\pm$ .010 \\
\hline
$\Sigma^-$  &
-1.133 & -0.361 & 0.017 &  0.444 & -0.156 & -1.190 & -1.160 $\pm$ .025 \\
\hline
$\Xi^0$  &
-1.467 & -0.040 & 0.000 &  0.349 & -0.175 & -1.333 & -1.250 $\pm$ .014 \\
\hline
$\Xi^-$  &
-0.467 & -0.219 & 0.000 &  0.028 & -0.031 & -0.689 & -0.651 $\pm$ .003 \\
\hline
$\Lambda$  &
-0.600 & -0.018 & 0.000 &  0.116 & -0.052 & -0.554 & -0.613 $\pm$ .004 \\
\hline
$\Sigma^0\rightarrow\Lambda$ &
 1.732 &  0.154 & 0.000 & -0.278 &  0.125 &  1.733 &  1.61  $\pm$ .08  \\
\hline
\end{tabular}
\caption{Different contributions to the octet baryon magnetic moments
in comparison to the experimental values. The first column ${\mu}_{imp}$
shows the
impulse result, to be compared with the total result (${\mu}_{tot}$) and the
experimental values in the last column. The second, third, forth and
fifth column contain respectively
the exchange current contributions due to the gluon, the pion (pion--pair
and pionic part), the confinement and the $\sigma$--exchange.
All quantities are given in n.m.\
${\mu}_N$= ${e\over 2M_P}$.
The last line contains the transition magnetic
moment for the decay $\Sigma^0\rightarrow\Lambda$.}
\label{table:mamo}
\end{center}
\end{table}

\begin{table}[b]
\begin{center}
\begin{tabular}{| l || r | r | r | r | r || c | c |} \hline
octet & $\langle r^2\rangle_{imp}$ & $\langle r^2\rangle_{gq\bar q}$ &
$\langle r^2\rangle_\pi$ &
  $\langle r^2\rangle_{conf}$ & $\langle r^2\rangle_\sigma$ &
  $\sqrt{\vert\langle r^2\rangle_{tot}\vert}$ &
  $r_{exp}$ \cite{Buc91} \\ \hline\hline
 p & 0.764 &  0.117  & 0.132 &  -0.385  & 0.043  & 0.819
   &  0.858 $\pm$ 0.056 \\ \hline
 n & 0.853 &  0.065  & 0.203 &  -0.430  & 0.048  & 0.860
   &  0.876 $\pm$ 0.070 \\ \hline
$\Sigma^+$  &
     0.702 &  0.116  & 0.003 &  -0.303  & 0.046  & 0.751  & \\ \hline
$\Sigma^-$  &
     0.683 &  0.164  &-0.008 &  -0.324  & 0.047  & 0.750  & \\ \hline
$\Xi^0$  &
     0.706 &  0.015  & 0.000 &  -0.206  & 0.047  & 0.750  & \\ \hline
$\Xi^-$  &
     0.370 &  0.162  & 0.000 &  -0.026  & 0.016  & 0.723  & \\ \hline
$\Lambda$  &
     0.763 &  0.017  & 0.000 &  -0.172  & 0.034  & 0.801  & \\ \hline
$\Sigma^0\rightarrow\Lambda$ &
     0.704 &  0.048  & 0.000 &  -0.206  & 0.026  & 0.756  & \\ \hline
\end{tabular}
\caption{Different contributions to the octet baryon magnetic radii, and
comparison with the experimentally available data. The notations are analogous
to the ones in table \ref{table:mamo}. All quantities are
given in [fm$^2$], in the last two columns in [fm].}
\label{table:mara}
\end{center}
\end{table}


\newpage
\begin{thebibliography}{99}
\bibitem{Rho79} Mesons in Nuclei II, edited by M.\ Rho and D.\ Wilkinson,
North Holland, New--York, 1979
\bibitem{Buc91}  A.\ Buchmann, E.\ Hernandez, and K.\ Yazaki,
Phys.Lett.{\bf B269} (1991), 35; \\
A.\ Buchmann, E.\ Hernandez, and K.\ Yazaki,
Nucl.Phys.{\bf A569} (1994), 661
\bibitem{Kar92} G.\ Karl, Phys.Rev.{\bf D45} (1992), 247
\bibitem{Clo79} F.E.\ Close, An Introduction to Quarks and Partons,
Academic Press, London, 1979
\bibitem{Buc89} A.\ Buchmann, Y.\ Yamauchi, and A.\ Faessler,
Nucl.Phys.{\bf A496} (1989), 621; \\
Phys.Lett.{\bf B225} (1989), 301;
Progr.Part.Nucl.Phys.{\bf 24}, ed.\ A.\ Faessler, p.\ 333
(Pergamon Press, Oxford, 1990)
\bibitem{Rev} for recent reviews see:
W.\ Lucha, F.F.\ Sch\"oberl, D.\ Gromes, Phys.Rep.{\bf 200} (1991), 127;
M.M.\ Giannini, Rep.Prog.Phys.{\bf 54} (1990), 453; \\
R.F.\ Alvarez--Estrada, F.\ Fernandez, J.L.\ Sanchez--Gomez and V.\ Vento,
Lecture notes in physics 259 (Springer, Berlin, 1986)
\bibitem{Shi84} K.\ Shimizu, Phys.Lett.{\bf B148} (1984), 418;
 D.\ Robson, Phys.Rev.{\bf D35} (1985), 1029; \\
 K.\ Maltman, Nucl.Phys.{\bf A446} (1985), 623; \\
 F.\ Fernandez and E.\ Oset, Nucl.Phys.{\bf A455} (1986), 720
\bibitem{Obu90} I.T.\ Obukhovsky and A.M.\ Kusainov, Phys.Lett.{\bf B238}
(1990), 142; \\
 F.\ Fernandez, A.\ Valcarce, U.\ Straub and A.\ Faessler,
J.Phys.{\bf G19} (1993), 2013
\bibitem{deR75} A.\ DeRujula, H.\ Georgi, and S.L.\ Glashow,
Phys.Rev.{\bf D12} (1975), 147
\bibitem{Buc95}  A.\ Buchmann, Czech.J.Phys.{\bf 45} (1995)
\bibitem{Val94} A.\ Valcarce, A.\ Buchmann, F.\ Fernandez,
and A.\ Faessler, Phys.Rev.{\bf C50} (1994), 2246; \\
Phys.Rev.{\bf C51} (1995), 1480
\bibitem{Vog90} U.\ Vogl, M.\ Lutz, S.\ Klimt, W.\ Weise,
Nucl.Phys.{\bf A516} (1990), 469
\bibitem{Isg78} N.\ Isgur and G.\ Karl, Phys.Rev.{\bf D18} (1978), 4187;
Phys.Rev.{\bf D19} (1979), 2653
\bibitem{Agu94} Particle Data Group, M.\ Aguilar--Benitez et al.,
Phys.Rev.{\bf D50} (1994), 1173
\end{thebibliography}
\end{document}